# Silicon Tetrafluoride on Io

Laura Schaefer and Bruce Fegley, Jr.

Planetary Chemistry Laboratory, Department of Earth and Planetary Sciences,
Washington University, St. Louis, MO 63130-4899
laura_s@levee.wustl.edu
bfegley@levee.wustl.edu

**ABSTRACT** Silicon tetrafluoride ($SiF_4$) is observed in terrestrial volcanic gases and is predicted to be the major F − bearing species in low temperature volcanic gases on Io (Schaefer and Fegley, 2005b). $SiF_4$ gas is also a potential indicator of silica − rich crust on Io. We used F/S ratios in terrestrial and extraterrestrial basalts, and gas/lava enrichment factors for F and S measured at terrestrial volcanoes to calculate equilibrium $SiF_4/SO_2$ ratios in volcanic gases on Io. We conclude that $SiF_4$ can be produced at levels comparable to the observed $NaCl/SO_2$ gas ratio. We also considered potential loss processes for $SiF_4$ in volcanic plumes and in Io's atmosphere including ion-molecule reactions, electron chemistry, photochemistry, reactions with the major atmospheric constituents, and condensation. Photochemical destruction ($t_{chem}$ ~266 days) and/or condensation as $Na_2SiF_6$ (s) appear to be the major sinks for $SiF_4$. We recommend searching for $SiF_4$ with infrared spectroscopy using its 9.7 µm band as done on Earth.

## INTRODUCTION

During our previous work on alkali halide chemistry on Io (Schaefer and Fegley, 2005b), we learned that silicon tetrafluoride is observed at several terrestrial volcanoes including Mount Iwodake, Vulcano, Mount Etna, and Popocatépetl (Francis et al., 1996; Love et al., 1998; Mori et al., 2002). Gaseous $SiF_4$ is detected on Earth by observing its 9.7 µm band with infrared (IR) absorption spectroscopy, which is used to measure $SiF_4/SO_2$ molar ratios. The observed $SiF_4/SO_2$ ratios range from 0.0004 at Mount Etna to 0.014 at Mount Iwodake. The $SiF_4/SO_2$ ratios increase as the volcanoes become more silicic (i.e., their lavas contain more $SiO_2$): Mount Etna (basaltic) 0.0004, Popocatépetl (andesitic/dacitic) 0.0011-0.011, Vulcano (rhyolitic) 0.0059, and Mount Iwodake (rhyolitic) 0.014. Generally, silicic lavas contain more F than mafic lavas (~1000 μg/g F in a rhyolite vs. ~250 μg/g in a basalt), consistent with the increase in the $SiF_4/SO_2$ ratio (Govindaraju, 1994). Keszthelyi and McEwen (1997) have suggested that Io has a silica-rich crust based on their magmatic differentiation model for Io. If this is true, then we should expect to find some silicic, fluorine-rich lavas on Io. Therefore it is plausible that $SiF_4$ could form in some Ionian volcanic gases generated from silica–rich magmatic reservoirs. If $SiF_4$ is observed on Io, it could help confirm the presence of silica-rich crustal regions on Io.

Our previous modeling (Schaefer and Fegley, 2005b) showed that $SiF_4$ can form in Ionian volcanic gases and is the most abundant fluorine-bearing compound at temperatures below ~900 K. In this work, we extend our study of $SiF_4$ in Ionian volcanic gases in two ways. First, we looked at the change in the $SiF_4/SO_2$ ratio of volcanic gases when the F/S ratio is varied. Second, we modeled several potential loss mechanisms for $SiF_4$ in volcanic plumes and Io's atmosphere including: ion-molecule chemistry, photochemistry, reactions with major atmospheric species, electron dissociation reactions, and condensation to determine if $SiF_4$ has a long enough residence time to be observable. Preliminary results are given by Fegley and Schaefer (2004).

**CHEMICAL EQUILIBRIUM CALCULATIONS**

We modeled chemical equilibria of O, S, Li, Na, K, Rb, Cs, F, Cl, Br, and I compounds as a function of temperature and pressure in an Ionian volcanic gas using the same methods as in our prior work (Schaefer and Fegley, 2005b). We took the nominal temperature and pressure for the Pele volcano as 1760±210 K and 0.01 bars as in our previous models. Our nominal elemental composition is listed in Table 1 of Schaefer and Fegley (2005b). We performed two sets of calculations (see Figures 5a and 5f of that paper) both with and without pure condensed $SiO_2$ (as quartz at low temperatures, and as cristobalite above the quartz→cristobalite phase transition temperature at 1079 K; Chase, 1998). For the first set of calculations, NaF (s,g) is the most abundant fluorine compound at temperatures and pressures found on Io. However, when pure silica is present, it reacts with fluorides in the gas to form $SiF_4$ (g), which becomes the most abundant fluorine-bearing compound at temperatures below about 900 K.

**RESULTS**

*Fluorine/Sulfur ratios in volcanic rocks*

Figure 1 shows the $SiF_4/SO_2$ ratios calculated for a pressure of 0.01 bars, the nominal pressure calculated for the Pele hot spot (Zolotov & Fegley, 2001), as a function of temperature for F/S ratios ranging from 0.003 to 900 × the CI chondritic value. The calculated values are compared to $SiF_4/SO_2$ ratios observed on Earth (Love et al., 1998; Francis et al., 1996; Mori et al., 2002). Calculations shown in Schaefer and Fegley (2005b) assumed a CI chondritic F/S atomic ratio of 0.00189 (Lodders, 2003). However, terrestrial volcanic rocks typically have F/S atomic ratios much larger than the CI chondritic value (F/S ~ 2 times the CI value in ultramafic rocks, 2-60 times the CI value in basaltic and andesitic rocks, and 10-200 times the CI value in rhyolitic rocks) (Govindaraju, 1994). By analogy, we might expect Ionian volcanic rocks to have F/S ratios larger than the CI chondritic value.

Here, we have chosen to use basalts as an analog to Ionian lavas because there is evidence for basaltic volcanism on Io (Kargel et al., 2003 and references therein); however, we are not excluding the possibility that there are also some ultramafic (Kargel et al., 2003) or silicic (Keszthelyi and McEwen, 1997) lavas present on Io, but rather limiting ourselves to the type of lava for which there is the most data, i.e. basalts. Even so, there is no reason to assume that Ionian lavas are strictly analogous to terrestrial basalts. Therefore, we also looked at basalts from other planetary bodies, such as lunar basalts, eucrites, and Martian basaltic meteorites. Eucrites are volatile-depleted basaltic meteorites believed to have originated from the asteroid 4 Vesta. For the seven eucrites that have been analyzed for both F and S, the range of F/S atomic ratios is 0.003-0.089, with a mean value of 0.04 (Kitts and Lodders, 1998). For basaltic Martian meteorites (shergottites), the F/S atomic ratio ranges from 0.027 – 0.061, with a mean of 0.039 (Lodders, 1998). For lunar basalts, we find a range of F/S atomic ratios of 0.053 – 0.123, with a mean value of 0.078 (Fegley & Swindle, 1993 and references therein). Combining the eucrites, Martian basalts, and lunar basalts, we get a mean F/S atomic ratio for volatile-depleted basalts of 0.05.

*Partitioning of fluorine and sulfur between lavas and volcanic gases*

We need to translate the F and S abundances in basalts into their abundances in a volcanic gas because they are probably degassed from lavas with different efficiencies. We do this using enrichment factors (E.F.$_i$), which measure the distribution of an element between two phases relative to a reference element. We need the gas (*g*) – lava (*l*) enrichment factor given by:

$$E.F._i = \frac{(C_{i,g}/C_{r,g})}{(C_{i,l}/C_{r,l})} \tag{1}$$

where *i* is the element being measured, *r* is the reference element, and *C* is the concentration of a given element in the gas or lava (Vié le Sage, 1983). Enrichment factors for fluorine and sulfur have been measured simultaneously at several terrestrial basaltic volcanoes (Andres et al., 1993; Crowe et al., 1987; Allard et al., 2000; Symonds et al., 1987, 1990).

Andres et al. (1993) measured enrichment factors for fluorine and sulfur of $10^4$ and $10^5$, respectively, at Mount Etna during a relatively quiescent period. The reference element used was Br; however, if we take the ratio of the F and S enrichment factors, the reference element cancels out and we are left with:

$$\left(\frac{C_F}{C_S}\right)_g = \frac{E.F.(F)}{E.F.(S)}\left(\frac{C_F}{C_S}\right)_l = 0.1\left(\frac{C_F}{C_S}\right)_l \tag{2}$$

If we assume a typical F/S ratio for a terrestrial basaltic lava of ~10 (Govindaraju, 1994), this gives us a F/S ratio of ~1 in the gas, which is ~530 times larger than the CI chondritic ratio. For a volatile-depleted basalt with F/S = 0.05, equation (2) gives a F/S gas ratio of $5\times10^{-3}$, which is ~2.6 × larger than the CI chondritic value. This data is listed in Table 1 along with similar calculations for data given in Crowe et al. (1987), Allard et al. (2000) and Symonds et al. (1987, 1990). Crowe et al. (1987) measured enrichment factors at Kilauea, a basaltic island volcano in Hawaii. Allard et al. (2000) measured F and S enrichment factors at Stromboli volcano, a K-rich basaltic volcano located in the Aeolian arc. Symonds et al. (1987) measured F and S enrichment factors at Merapi, an andesitic stratovolcano with high temperature fumaroles in Indonesia. Symonds et al. (1990) measured F and S enrichment factors for Augustine Volcano, an andesitic/dacitic volcano with significant seawater contamination. Apparently no gas/lava enrichment factors have been measured at rhyolitic or ultramafic volcanos.

Before we apply data for terrestrial enrichment factors to Ionian lavas, we should mention that enrichment factors can be affected by a variety of different conditions including, but not limited to: temperature, silica content, sulfur content, and water content. Therefore, no two basaltic volcanoes will have identical enrichment factors for each element, as can be seen in Table 1, and the enrichment factor for an element may vary at a single volcano with time due to changes in conditions. The same should also be true of Ionian volcanoes, i.e., it is doubtful that all Ionian volcanoes have identical lavas under identical conditions, so a spread of values for enrichment factors is to be expected. However, the most important difference between terrestrial and Ionian volcanoes is water; Io is very dry so there should be significantly less water in Ionian lavas than in terrestrial lavas, which may affect how elements are partitioned between gas and lava. We therefore discarded enrichment factors measured at Augustine volcano, which has a very high water content (Symonds et al., 1990).

Finally, we combined the data for terrestrial basalts from Andres et al. (1993), Allard et al. (2000), and Crowe et al. (1987) and the data for Merapi volcano from Symonds et al. (1987), which gives us a range of F/S gas ratios of 0.5 – 900 × the CI chondritic value for typical terrestrial basaltic lavas, and 0.003 – 4.46 × the CI chondritic value for a volatile-depleted basalt. From these calculations, we estimate a possible range of F/S gas ratios of 0.003 – 900 × the CI chondritic value for all basaltic lavas.

We can compare these values to the atmosphere of Venus, which is produced primarily through volcanic out-gassing. The surface of Venus is mostly basaltic, so the F/S atmospheric ratio should give us a value comparable to those we calculated above for basaltic lavas (Fegley, 2004). In Venus' atmosphere, sulfur is present primarily as $SO_2$, with minor amounts present in OCS, $H_2S$, SO, and $S_x$, where $x$ =1-8. Fluorine is found entirely in HF. Using abundances of these gases from Lodders and Fegley (1998), we calculate a F/S gas ratio of $\sim 3 \times 10^{-5}$. This value is ~0.02 × the CI chondritic value, which falls within the range of gas ratios we calculated for a typical volatile-depleted basalt.

*$SiF_4/SO_2$ ratios in Ionian volcanic gases*

Figure 1 shows the $SiF_4/SO_2$ ratio computed using F/S gas ratios spanning the calculated range discussed above, including: 0.003 × CI chondritic (~$5.67 \times 10^{-6}$), 0.5 × CI chondritic ($9.45 \times 10^{-4}$), CI chondritic (0.00189 – nominal value), 4 × CI chondritic (0.00756), 50 × CI chondritic (0.0945), and 900 × CI chondritic (~1.701). The calculated SiF$_4$/SO$_2$ ratio for a F/S ratio equal to 0.003 × CI chondritic is ~2×10$^{-6}$ below ~620 K, and drops off to negligible amounts at higher temperatures. These values are smaller than any observed abundances on the Earth. The calculated SiF$_4$/SO$_2$ ratio for a F/S ratio equal to 0.5 × CI chondritic is ~3.5×10$^{-4}$ below ~550 K, which is slightly smaller than the value observed at Mount Etna (basaltic). The SiF$_4$/SO$_2$ ratio then drops slightly to 3.1×10$^{-4}$ above 550 K, due to appearance of LiF (s), and falls off to zero near 900 K, due to appearance of NaF (s). The similar kinks and structure of the curves for higher F/S ratios are also due to the presence of LiF (s) and NaF (s), which consume fluorine. The SiF$_4$/SO$_2$ ratio calculated for a CI chondritic F/S ratio is ~7×10$^{-4}$ and falls between the observed values for Mount Etna (basaltic) and the lowest observed abundance at Popocatépetl (andesitic/dacitic). The calculated SiF$_4$/SO$_2$ ratio for an F/S ratio of 4 × CI chondritic is ~3×10$^{-3}$ at temperatures below 900 K, which falls between the range of values observed for Popocatépetl. An F/S ratio of 10× CI chondritic agrees well with the observed values for Popocatépetl, Mount Iwodake, and Vulcano. The calculated SiF$_4$/SO$_2$ ratio for an F/S ratio of 50 × CI chondritic is ~0.03, which is larger than any value observed on Earth. The calculated SiF$_4$/SO$_2$ ratio for an F/S ratio of 900 × CI chondritic is an essentially constant value of ~0.27.

In comparison to our predicted SiF$_4$ abundances, we note that NaCl (g) was observed at similar abundance levels through Earth-based millimeter wave spectroscopy by Lellouch et al. (2003). They determined a disk-averaged NaCl/SO$_2$ ratio of ~0.3%, but favored localized dense atmospheres of NaCl near active volcanic vents, where abundances of NaCl could vary from ~0.2-10% of SO$_2$. The disk-averaged abundance of NaCl (~0.3% of SO$_2$) is almost identical to our SiF$_4$/SO$_2$ abundances for an F/S ratio equal to 4 times the CI chondritic value. While some lavas on Io may give smaller SiF$_4$ abundances as did the volatile-depleted basalts, there may also be lavas on Io which are more silica-rich, as predicted by Keszthelyi and McEwen (1997), which would generate larger SiF$_4$ abundances. Therefore, SiF$_4$ could have comparable abundances to compounds already observed at Io, so it may be observable, provided it has a long enough atmospheric residence time.

**DESTRUCTION OF SIF$_4$**

We investigated a number of loss mechanisms for SiF$_4$ in volcanic plumes and Io's atmosphere to determine if SiF$_4$ will be present in Io's atmosphere long enough to be observed. We looked at ion-molecule reactions analogous to those that destroy perfluorocarbons (e.g., CF$_4$), chlorofluorocarbons (e.g., CF$_2$Cl$_2$), and sulfur fluorides (e.g., SF$_6$) in the Earth's atmosphere. We also calculated the photochemical lifetime of SiF$_4$ and looked at reactions of SiF$_4$ with major atmospheric constituents such as O, S, SO$_2$, Na, K, and Cl, and at electron dissociation reactions. A final possible destruction mechanism for SiF$_4$ is condensation either as SiF$_4$ (ice) or as Na$_2$SiF$_6$ (s). We discuss each of these possible sinks below.

*Ion-molecule reactions* There is very little kinetic data available for SiF$_4$, but we do have kinetic data for similar compounds such as CF$_4$ and C$_2$F$_6$, which are greenhouse gases found in the Earth's atmosphere. Silicon tetrafluoride could be destroyed in a similar manner. Carbon tetrafluoride and C$_2$F$_6$ are very long-lived compounds in the Earth's atmosphere, and they are destroyed primarily through high-temperature combustion in power plants and automobiles (Cicerone, 1979; Ravishankara et al., 1993). Atmospheric destruction processes for these compounds take place within the mesosphere and thermosphere (Cicerone, 1979; Ravishankara et al., 1993). Morris et al. (1995a, b) studied ion-molecule reactions involving various fluorocarbon and chlorofluorocarbon compounds including CF$_4$ and C$_2$F$_6$. They found that CF$_4$ and C$_2$F$_6$ do not react with NO$^+$, H$_3$O$^+$, NO$_3^-$, CO$_3^-$, O$_2^+$, and O$^-$. However, both fluorocarbons react with O$^+$ ions (Morris et al. 1995a, b):

$$CF_4 + O^+ = CF_3^+ + FO \quad (3)$$

$$k_3 = 1.4 \times 10^{-9} \text{ cm}^3 \text{ s}^{-1} \qquad T = 300 \text{ K}. \quad (4)$$

$$C_2F_6 + O^+ = CF_3^+ + CF_3O \quad (5)$$

$$k_5 = 1.1 \times 10^{-9} \text{ cm}^3 \text{ s}^{-1} \qquad T = 300 \text{ K}. \quad (6)$$

The estimated minimum lifetimes of CF$_4$ and C$_2$F$_6$ on Earth from reactions (3) and (5) are 330,000 and 420,000 years, respectively (Morris et al. 1995b). These reactions are the primary atmospheric loss mechanisms for CF$_4$ and C$_2$F$_6$ on Earth; neutral reactions and photolysis give minimum lifetimes for both compounds of >1 million years (Morris et al. 1995b). Chlorofluorocarbons are also destroyed by ion-molecule reactions on Earth;

however, these reactions attack the C-Cl bond, which is weaker than the C-F bond and is therefore more readily destroyed. If $SiF_4$ reacts with $O^+$ ions with the same rate constant as $CF_4$ and $C_2F_6$, then we find two different chemical lifetimes for $SiF_4$ destruction:

$$t_{chem} [SiF_4] = 1/ k_3 [O^+] \qquad (7)$$

$$t_{chem} [SiF_4] = 1/ k_5 [O^+] \qquad (8)$$

where $[O^+]$ is the concentration of $O^+$ ions in $cm^{-3}$. At Io, the major source of $O^+$ ions is through photochemistry. Moses et al. (2002) give an $O^+$ abundance of ~0.1 $cm^{-3}$ at 0 km altitude increasing to ~6 $cm^{-3}$ at 150-500 km. Summers and Strobel (1996) give a slightly lower $O^+$ abundance of ~0.01 $cm^{-3}$ at 0 km increasing to ~3 $cm^{-3}$ at 150-400 km. Substituting these $O^+$ number densities into equations (7) and (8) gives a $t_{chem}$ for $SiF_4$ of 4 – 2265 years from equation (7) and 5 – 2900 years from equation (8), where shorter lifetimes correspond to higher altitudes where $O^+$ is more abundant.

*Photodissociation* $SiF_4$ is photochemically destroyed by extreme ultraviolet light (EUV) with a long wavelength limit of ~170 nm. We calculated the photodissociation constant ($J_1$) for $SiF_4$ at zero-optical depth using the UV absorption cross section from Suto et al. (1987). At 1 AU, the $J_1$ value for $SiF_4$ is $1 \times 10^{-6}$ $s^{-1}$, which gives a photochemical lifetime of ~10 days at Earth. For comparison, the $J_1$ value for $CF_4$ at zero-optical depth is ~6.8 × $10^{-7}$ $s^{-1}$, which gives a photochemical lifetime of ~17 days; however, absorption of light by $O_2$ and $N_2$ in the Earth's atmosphere drastically increases the photochemical lifetime of $CF_4$ (Cicerone, 1979). For $SiF_4$, when scaled to 5.2 AU, the $J_1$ value gives a photochemical lifetime at zero-optical depth of ~266 days for $SiF_4$ in Io's atmosphere. Therefore if photochemistry is the only destruction mechanism, $SiF_4$ is fairly long-lived at Io and should be present in the atmosphere several months after it is generated in a volcanic eruption.

*Atmospheric reactions* Once out-gassed, it is possible that $SiF_4$ is destroyed by chemical reactions with other gases in the volcanic plume or in Io's atmosphere. We therefore looked at reactions involving $SiF_4$ and the major atmospheric constituents (O, S, $SO_2$, Na, K, Cl). There are no reaction rates available for these reactions. However, all of these reactions are highly endothermic:

$$SiF_4 + O = SiF_3 + FO \qquad \Delta_9 H°_{298} = +477 \text{ kJ mole}^{-1} \qquad (9)$$

$$SiF_4 + S = SiF_3 + SF \qquad \Delta_{10} H°_{298} = +356 \text{ kJ mole}^{-1} \qquad (10)$$

$$SiF_4 + SO_2 = SiF_2 + SO_2F_2 \quad \Delta_{11}H°_{298} = +559 \text{ kJ mole}^{-1} \quad (11)$$

$$SiF_4 + Cl = SiF_3Cl + F \quad \Delta_{12}H°_{298} = +255 \text{ kJ mole}^{-1} \quad (12)$$

$$SiF_4 + Na = SiF_3 + NaF \quad \Delta_{13}H°_{298} = +215 \text{ kJ mole}^{-1} \quad (13)$$

$$SiF_4 + K = SiF_3 + KF \quad \Delta_{14}H°_{298} = +200 \text{ kJ mole}^{-1} \quad (14)$$

Reactions (13) and (14) are the least endothermic and should be faster than the other reactions. However, scattering experiments indicate that collisions between $SiF_4$ and Na are non-reactive (Düren et al., 1998). It is therefore unlikely that $SiF_4$ will be destroyed by reactions with the major gases in volcanic plumes or Io's atmosphere.

*Electron Dissociation Reactions* Electron dissociation reactions such as

$$SiF_4 + e^- = SiF_3^+ + F \quad (15)$$

may also destroy $SiF_4$ in Io's atmosphere. However, electron dissociation reactions are relatively unimportant for analogous compounds such as $CF_4$, $C_2F_6$ and $SF_6$ in the Earth's atmosphere due to the small global abundance of these compounds at altitudes where electrons can penetrate the atmosphere (Cicerone, 1979). Electron-dissociation reactions with $SiF_4$ have a minimum threshold energy of ~10.8 eV (Basner et al., 2001). At Io, the major sources of electrons are the Io plasma torus (IPT) and the ionosphere. The IPT has an average electron temperature of ~5 eV, with a much smaller population of more energetic electrons, and the electrons in the ionosphere are cooler than torus electrons due to collisions with neutrals in Io's atmosphere (Saur et al., 2002; Summers and Strobel, 1996). The electron abundance decreases with decreasing altitude in Io's atmosphere due to attenuation by interaction with $SO_2$ (Smyth and Wong, 2004). Thus, $SiF_4$ destruction by electron dissociation reactions is probably insignificant on Io.

*Condensation* Another possible loss mechanism for $SiF_4$ is condensation. Figure 2 shows vapor pressure curves for $SO_2$, $CO_2$, and $SiF_4$ (Lyman and Noda, 2001). The shaded region on the graph represents the average surface conditions of Io. The average surface temperature of Io ranges from 90-130 K (Rathbun et al., 2004), and the average atmospheric pressure is 1-10 nanobars (Lellouch, 1996). In order for an ice to be stable on Io's surface, the vapor pressure must fall within or below the average range of atmospheric pressures for the average surface temperatures. As can be seen, while the vapor pressure curve of $SO_2$ overlaps very well with the average surface conditions as expected, $CO_2$ ice barely overlaps, and $SiF_4$ ice not at all. Whereas we might expect to find CO$_2$ ice condensed in some slightly colder than average areas of the surface (Schaefer and Fegley, 2005a), SiF$_4$ would require significantly colder temperatures. Therefore, condensation as SiF$_4$ ice is probably not an important loss process for SiF$_4$ (g).

On the other hand, SiF$_4$ may also condense by reaction with NaF to form Na$_2$SiF$_6$:

$$\text{SiF}_4 \text{ (g)} + 2 \text{ NaF (s,g)} = \text{Na}_2\text{SiF}_6 \text{ (s)} \tag{16}$$

Na$_2$SiF$_6$ (s) forms from slags used in steelmaking (Kashiwaya and Cramb, 1998). We did chemical equilibrium calculations using data for Na$_2$SiF$_6$ from Stull et al. (1970) and Chiotti (1981) and found that reaction (16) is only important at temperatures much lower than magmatic temperatures. For example, Na$_2$SiF$_6$ condensation temperatures are 105 K (F/S ratio = 0.1 × CI), 135 K (F/S = CI chondritic), and 300 K (F/S = 50 × CI). However, the SiF$_4$ vapor pressure over Na$_2$SiF$_6$ (Caillat, 1945) extrapolates to 10$^{-50}$ bar at 110 K. This is significantly less than Io's atmospheric pressure and Na$_2$SiF$_6$ (s) should be stable on Io's surface. Thus, although it is unstable at active vents, Na$_2$SiF$_6$ may condense on the surface away from high temperature volcanic vents.

## SUMMARY

We used analyses of fluorine and sulfur in terrestrial and extraterrestrial basalts and gas/lava enrichment factors for F and S at terrestrial volcanoes to estimate F/S atomic ratios in Ionian volcanic gases. We derived F/S gas ratios of 0.5 – 900 × the CI chondritic value from F and S abundances in terrestrial basaltic lavas, and 0.003 – 4.46 × the CI chondritic value from F and S abundances in volatile-depleted basalts from the Moon, eucrite meteorites, and the SNC meteorite parent body (presumably Mars). We calculated chemical equilibrium SiF$_4$/SO$_2$ ratios in Ionian volcanic gases using F/S ratios of 0.003 – 900 × the CI chondritic value. Larger F/S ratios of 50 – 900 times CI chondritic give larger SiF$_4$/SO$_2$ ratios than at terrestrial volcanoes, which seems unlikely. Fluorine to sulfur ratios < 0.1 × the CI chondritic value will give proportionally lower SiF$_4$/SO$_2$ ratios on Io. Our calculated SiF$_4$/SO$_2$ ratios for Ionian volcanic gases are comparable to those in terrestrial volcanoes and to the observed NaCl/SO$_2$ ratio in Io's atmosphere (Love et al., 1998; Mori et al., 2002; Francis et al., 1996; Lellouch et al. 2003). We examined several possible sinks for SiF$_4$ in volcanic plumes and Io's atmosphere. Photochemical destruction by EUV sunlight (t$_{chem}$ ~266 days) and/or condensation as Na$_2$SiF$_6$ (s) appear to be the major sinks for $SiF_4$. We suggest that $SiF_4$ will likely remain in Io's atmosphere long enough to be observed after it is produced in volcanic gases. Based upon its predicted chemical equilibrium abundance, long lifetime, and IR observations (at 9.7 µm) distinguishing $SiF_4$ from $SO_2$ in terrestrial volcanic gases, we recommend searching for $SiF_4$ (g) at Io using one or more of its major IR bands at 4.86, 5.47, 7.73, 8.40, 8.61, 9.39, 9.71, and 25.7 µm.

**ACKNOWLEDGMENTS**

This work was supported by the NASA Outer Planets Research Program. We would like to thank L. Keszthelyi and M. Summers for constructive reviews.

**Table 1**.
Calculated F/S atomic gas ratios for terrestrial and volatile-depleted basalts.

| Volcano | EF(F)/EF(S) | $(F/S)_l$ | $(F/S)_g$ | $(F/S)_g \times CI$ | Reference |
|---|---|---|---|---|---|
| Mount Etna | 0.1 | 10<br>0.05 | 1<br>$5\times10^{-3}$ | 530<br>2.6 | Andres et al. (1993) |
| Kilauea | $10^{-4} - 3.2\times10^{-3}$ | 10<br>0.05 | $0.001 - 0.032$<br>$5\times10^{-6} - 2\times10^{-4}$ | $0.5 - 17$<br>$0.003 - 0.08$ | Crowe et al. (1987) |
| Stromboli | $3.1\times10^{-3}$ | 10<br>0.05 | 0.03<br>$1.6\times10^{-4}$ | 16<br>0.08 | Allard et al. (2000) |
| Merapi | 0.17 | 10<br>0.05 | 1.7<br>$8.4 \times 10^{-3}$ | 900<br>4.46 | Symonds et al. (1987) |
| Augustine | $1 - 6300$ | 10<br>0.05 | 10-63000<br>$0.05 - 315$ | $5\times10^3 - 3\times10^7$<br>$26 - 1.7\times10^5$ | Symonds et al. (1990) |

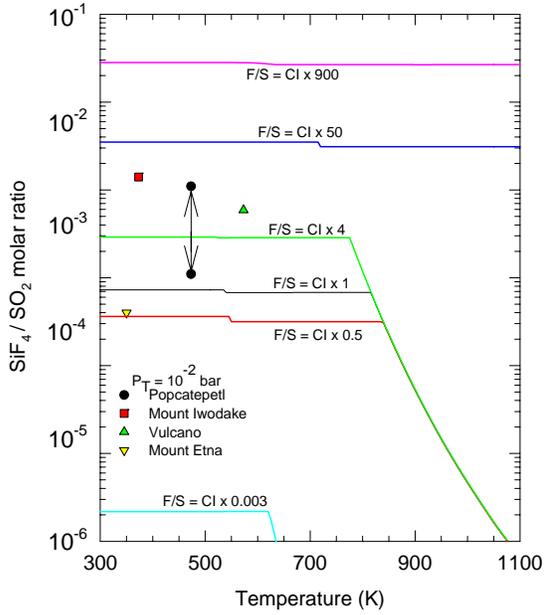

**Figure 1.** SiF$_4$/SO$_2$ ratio in an Ionian volcanic gas at a constant P = 0.01 bars as a function of temperature. The lines show calculations performed for F/S ratios of CI chondritic × 0.003, CI chondritic × 0.5, CI chondritic × 1, CI chondritic × 4, CI chondritic × 50, and CI chondritic × 900. The points show measured values for the SiF$_4$/SO$_2$ ratios in some terrestrial volcanic gases. References are given in the text.

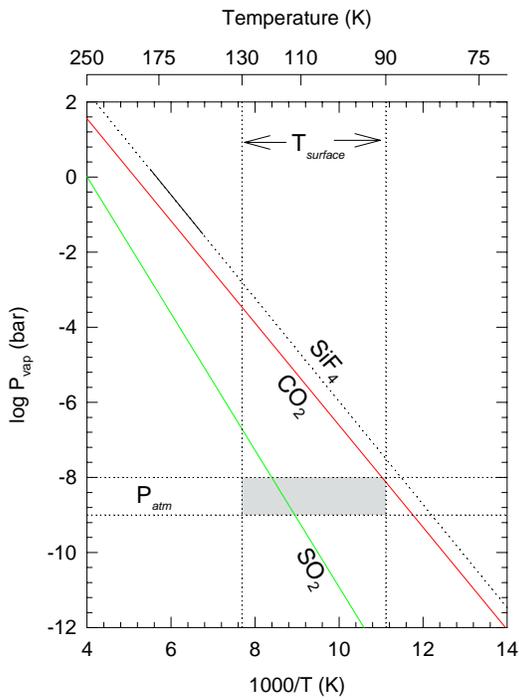

**Figure 2.** Vapor pressures of SiF$_4$, CO$_2$, and SO$_2$ ices. Solid lines indicate measured data, and dashed lines indicate extrapolated values. The range of Io's average atmospheric pressures (1-10 nbar, Lellouch, 1996) and surface temperatures (90-130 K, Rathbun et al., 2004) are shown as dashed lines and the shaded region illustrates the possible surface conditions.